\documentclass[twocolumn,prl,showkeys]{revtex4}%
\usepackage{graphicx}%
\usepackage{amsmath}%
\usepackage{amsfonts}%
\usepackage{amssymb}

\def\k{{ {\bf k} }}

\def\q{{ {\bf q} }}
\def\Q{{ {\bf Q} }}

\def\w{{\omega}}
\def\a{{\alpha}}
\def\b{{\beta}}

\begin{document}
\title{
Impurity Effects in Sign-Reversing Fully Gapped Superconductors:\\
Analysis of FeAs Superconductors
}
\author{Yuko \textsc{Senga} and Hiroshi \textsc{Kontani}}
\date{\today }

\begin{abstract}
To understand the impurity effect on $T_{\rm c}$
in FeAs superconductors, we analyze a simple two-band BCS model
with a repulsive interband interaction.
The realized fully gapped superconducting state with sign reversal,
which is predicted by spin fluctuation theories in this compound,
is suppressed by impurities due to the interband hopping of Cooper pairs,
if the interband impurity scattering $I'$ is equal to the intraband one $I$.
When $|I'/I|<1$, in high contrast, 
$T_{\rm c}$ is almost unchanged by strong impurity scattering
since interband scattering is almost prohibited by the 
multiple scattering effect.
Since $|I'/I|\sim0.5$ is expected,
the robustness of superconductivity against impurities
in FeAs superconductors is naturally understood in terms of 
the sign-reversing fully gapped state.
%predicted by spin fluctuation theories

\end{abstract}


%\draft

\address{
Department of Physics, Nagoya University, Furo-cho, Nagoya 464-8602, Japan.
}

\sloppy

\maketitle

%%%%%%%%%%%%%%%%%%
%Introduction
%%%%%%%%%%%%%%%%%%

\begin{figure}[!htb]
\includegraphics[width=.6\linewidth]{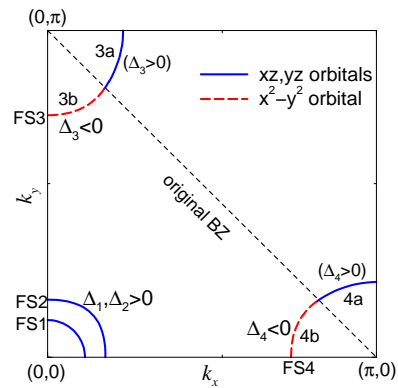}
\caption{
Fermi surfaces (FSs) in FeAs superconductors
in the unfolded Brillouin zone; see ref. \cite{Kuroki}.
}
\label{fig:FS}
\end{figure}

After the discovery of the superconductor La(O,F)FeAs ($T_{\rm c}=26$ K)
 \cite{Hosono},
high-$T_{\rm c}$ superconductors with FeAs layers have been studied intensively
 \cite{Ren,XHChen,GFChen,Eisaki}.
The ground state of the undoped compound is the spin density wave (SDW) 
state, where the ordered magnetic moment is $\sim0.3 \ \mu_{\rm B}$
and the ordering vector is $\Q\approx (\pi,0)$ or $(0,\pi)$
 \cite{Moss,Luetkens,neutron-SDW,neutron-noRP}.
The superconducting state is realized next to the SDW state by carrier doping
 \cite{Luetkens,Liu}
According to NMR study, the singlet superconducting state is
realized in FeAs
 \cite{Kawabata1,Kawabata2,Zheng}.
In the first-principle band calculations \cite{Singh,Ishibashi}, 
the Fermi surfaces in FeAs are composed of
two hole-like Fermi pockets around the $\Gamma=(0,0)$ point 
(FS1 and FS2 in Fig. \ref{fig:FS})
and  two electron-like Fermi pockets around M$=(\pi,0),(0,\pi)$ points
(FS3 and FS4 in Fig. \ref{fig:FS}).

Theoretically, there are several possible pairing states:
According to the random phase approximation (RPA)
based on a realistic five-orbital tight-binding model \cite{Kuroki},
the nesting between the hole and electron pockets
gives rise to the strong antiferromagnetic (AF) fluctuations 
with $\Q\approx (\pi,0)$, which is consistent with experimental results.
Then, a fully gapped $s$-wave state with sign reversal is expected to emerge
since AF fluctuations works as the repulsion interaction
between hole and electron pockets
 \cite{Kuroki,Mazin,RG,Nomura,Yanagi,Tesanovic}.
Although AF fluctuations due to the nesting between two electron pockets 
[$\q\sim(\pi,\pi/2)$]
can induce the $d_{x^2\mbox{-}y^2}$-wave state with line nodes on the 
hole pockets, the obtained $T_{\rm c}$ is rather low
\cite{Kuroki,RG,Nomura,Yanagi}.
On the other hand, the conventional $s$-wave state without sign reversal
will be realized if the charge fluctuations or electron-phonon 
interactions are strong.

Experimentally, a fully gapped superconducting state has been determined
by recent penetration depth measurement \cite{Matsuda}, 
angle-resolved photoemission spectroscopy (ARPES)
 \cite{ARPES1,ARPES2,ARPES3},
and specific heat measurement \cite{Mu}.
The resonance peak observed by inelastic neutron measurement below 
$T_{\rm c}$ supports the sign-reversing superconducting gap 
mediated by AF fluctuations \cite{Scalapino,neutron-RP}.
Anomalous transport phenomena (such as Hall coefficient and 
Nernst signal) in the normal state in FeAs,
which are similar to those observed in high-$T_{\rm c}$ cuprates
and Ce$M$In$_5$ ($M$=Co, Rh, Ir) \cite{Nakajima},
also indicate the existence of strong AF fluctuations
 \cite{Liu,Nernst,Kawabata2,Kontani-review}.
On the other hand, impurity effect on $T_{\rm c}$ due to Co, Ni, or Zn 
substitution for Fe sites is very small or absent
 \cite{Kawabata1,Kawabata2,Co,Ni,Zn},
which is decisive for determining the pairing symmetry.
For example, this result 
clearly rules out the possibility of line-node superconductivity.
This result may also eliminate the theoretically predicted sign-reversing 
$s$-wave states, since the Cooper pair is destroyed by
the interband scattering induced by impurities;
$(\k\uparrow,-\k\downarrow)_{\rm band1} 
\rightarrow (\k'\uparrow,-\k'\downarrow)_{\rm band2}$.
In this manner, a study of the impurity effect on FeAs superconductors
offers us significant information on the pairing symmetry,
and therefore reliable theoretical analyses are highly required.

In this letter, we study a simple two-band BCS model
to investigate the impurity effect on FeAs superconductors.
%using the $T$-matrix approximation.
Therein, the sign-reversing pairing state is realized
if we introduce the interband repulsive interaction to 
describe the effective interaction due to AF fluctuations.
When the interband impurity scattering potential $I'$ is equal to 
the intraband one $I$, 
the reduction in $T_{\rm c}$ per impurity concentration $n_{\rm imp}$,
$-\Delta T_{\rm c}/n_{\rm imp}$, is prominent as 
in non-$s$-wave superconductors.
However, $x=|I'/I|$ is smaller than 1 in this compound
since hole and electron pockets are not composed of the same $d$-orbitals.
In this case, $-\Delta T_{\rm c}/n_{\rm imp}$ becomes very small;
in particular, it approaches {\it zero} in the unitary regime 
($I/W_{\rm band}\gg1$).
Therefore, the experimental absence of the impurity effect on $T_{\rm c}$ 
in FeAs is well understood in terms of the sign-reversing $s$-wave state
proposed in refs. \cite{Kuroki} and \cite{Mazin}.

Here, we analyze the Eliashberg gap equation for the two-band BCS model.
To concentrate on studying the impurity effect,
we neglect the mass-enhancement factor and the quasiparticle damping 
due to electron-electron interaction for simplicity, 
both of which are given by the normal self-energy.
In the absence of impurities, the linearized gap equation 
at $T_{\rm c}$ is given by \cite{Allen}:
\begin{eqnarray}
%Z_i(p_n)&=& 1+\frac{\pi T_{\rm c}}{|p_n|}\sum_{m;j=\a,\b}'g_{ij}N_js_n s_m,
% \nonumber \\
%Z_i(p_n)
\Delta_i(p_n) &=& -\pi T_{\rm c}
 \sum_{j=\a,\b}g_{ij}N_j {\sum_{m}}'\Delta_j(p_m)/|p_m| ,
 \label{eqn:Eliash1}
\end{eqnarray}
where $p_n,p_m$ are fermion Matsubara frequencies,
and $i,j\ (=\a,\b)$ are band indices.
$N_i$ is the density of states at the Fermi level, and
$g_{ij}$ is the effective interaction between band $i$ and band $j$.
$\sum_m'\equiv \sum_m\theta(\w_c-|p_m|)$,
where $\w_c$ is the characteristic energy scale of the effective interaction.
$\Delta_i(p_n)$ is the gap function, which is independent of 
$p_n$ for $|p_n|<\w_c$ in the absence of impurities.
%$Z_i(p_n)\equiv 1-(\Sigma^n(p_n)-\Sigma^n(-p_n))/ip_n =1$
%if we neglect the normal self-energy $\Sigma^n(p_n)$ 
%in the BCS approximation.
Hereafter, we assume $g_{\a\b}\equiv g>0$
to realize the sign-reversing $s$-wave gap ($\Delta_\a\Delta_\b<0$),
and put $g_{\a\a}=g_{\b\b}=0$ for simplicity since they will be
much smaller than $g$ in FeAs.
After the standard analysis \cite{Allen},
the transition temperature without impurities is obtained as
$T_{\rm c}^0= 1.13\w_c \exp(-1/g\sqrt{N_\a N_\b})$.

%%%%%%%%%%%%%%%%%%%%%%%%%%%%%%%%%
\begin{figure}[!htb]
\includegraphics[width=.99\linewidth]{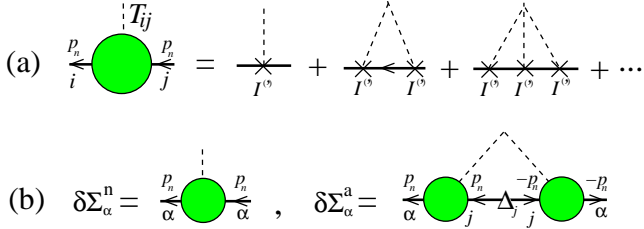}
\caption{
(a) $T$-matrix in the normal state.
(b) Impurity-induced normal self-energy ($\delta\Sigma^n_\a$)
and the linealized anomalous self-energy ($\delta\Sigma^a_\a$)
in the $T$-matrix approximation.
}
\label{fig:diagram}
\end{figure}
%%%%%%%%%%%%%%%%%%%%%%%%%%%%%%%%%

Hereafter, we study the nonmagnetic impurity effect 
in the two-band BCS model with $g_{\a\b}>0$.
%There are several pioneering works for $g_{\a\b}<0$ to analyze MgB$_2$
%\cite{Mg1,Mg2},
%which is a multiband $s$-wave superconductor due to 
%electron-phonon interaction.
In similar models, the effect of interband impurity scattering 
in the NMR relaxation ratio has been analyzed
using the Born approximation \cite{Chubukov}
and $T$-matrix approximation for $I=I'=\infty$ \cite{Parker}.
%From now on, we study the nonmagnetic impurity effect 
%in the two-band BCS model.
%There are several pioneering works for $g_{\a\b}<0$ to analyze MgB$_2$
%\cite{Mg1,Mg2},
%which is a multiband $s$-wave superconductor due to 
%electron-phonon interaction.
Very recently, Bang et al. have reported a sizable reduction in 
$T_{\rm c}$ by a strong impurity potential with $I=I'$ \cite{Bang}.
In contrast to their result, 
we will show below that $T_{\rm c}$ is almost unchanged 
when $I'/I<1$.

Hereafter, we use the $T$-matrix approximation, which 
gives the exact result for $n_{\rm imp}\ll1$
for any strength of $I, I'$.
We can assume that $I,I'\ge0$ without losing generality.
Using the local normal Green function 
$g_i(p_n)=-i\pi N_i s_n$ ($s_n\equiv {\rm sgn}(p_n)$) \cite{Allen},
the $T$-matrix in the normal state is given by
\begin{eqnarray}
T_{ij}(p_n)= I_{ij}+\sum_{l=\a,\b} I_{il}g_l(p_n)T_{lj}(p_n),
 \label{eqn:T}
\end{eqnarray}
where $I_{ij}=I\delta_{i,j}+I'(1-\delta_{i,j})$.
%$ is the impurity scattering potential between $i$-band and $j$-band.
Its diagrammatic expression is shown in Fig. \ref{fig:diagram} (a).
Except at $x=1$, the solution of eq. (\ref{eqn:T}) behaves as
$T_{\a\a}\sim -is_n/\pi N_\a +O(I^{-1})$ and
$T_{\a\b}\sim O(I^{-1})$ in the unitary limit.
This fact means that the {\it the superconducting state is unaffected
by a strong impurity potential since the interband scattering is prohibited
as a result of multiple scattering}. % \cite{Hirashima}.

Using the $T$-matrix, the impurity-induced normal and 
anomalous self-energies %of order $O(\Delta)$ 
are respectively given as 
\begin{eqnarray}
\delta \Sigma^n_i(p_n)&=& n_{\rm imp}T_{ii}(p_n),
 \\
\delta \Sigma^a_i(p_n)&=& n_{\rm imp} \sum_{l=\a,\b}
T_{il}(p_n)f_l(p_n)T_{li}(-p_n) ,
\end{eqnarray}
where $f_i(p_n) \equiv \pi N_i \Delta_i(p_n)/|p_n|$
is the local anomalous Green function at $T_{\rm c}$.
Their expressions are shown in Fig. \ref{fig:diagram} (b).
Then, the gap equation at $T_{\rm c}$ is given by \cite{Allen}
\begin{eqnarray}
Z_i(p_n)\Delta_i(p_n) 
&=& -\pi T_{\rm c}\sum_{j=\a,\b}g_{ij}N_j {\sum_m}'\Delta_j(p_m)/|p_m|
 \nonumber \\
& &+\delta\Sigma_i^a(p_n) ,
 \label{eqn:Eliash2}
\end{eqnarray}
where $Z_i(p_n)\equiv 1-(\delta\Sigma_i^n(p_n)-\delta\Sigma_i^n(-p_n))/2ip_n
= 1+\gamma_i/|p_n|$, and
$\gamma_i= -{\rm Im}\Sigma_i^n(p_n) \cdot s_n \ (>0)$ 
is the quasiparticle damping rate due to impurity scattering.

Now, we analyze the impurity effect on $T_{\rm c}$
in the case of $N_\a=N_\b \equiv N$ as the first step.
In this case, the relationships $\gamma_\a=\gamma_\b \equiv\gamma$,
$\Delta_\a(p_n)=-\Delta_\b(p_n) \equiv\Delta(p_n)$, and
$\delta\Sigma^a_\a=-\delta\Sigma^a_\b \equiv \delta\Sigma^a$ are satisfied.
They are obtained as
\begin{eqnarray}
& &\gamma = n_{\rm imp}\pi N[(I^2+{I'}^2)+\pi^2N^2 (I^2-{I'}^2)^2]/A ,
 \label{eqn:gamma} \\
& &\delta\Sigma^a(p_n) = n_{\rm imp}\Delta(p_n)\pi N [(I^2-{I'}^2)
 \nonumber \\
& & \ \ \ \ \ \ \ \ \ \ \ \ \ \ 
+\pi^2N^2 (I^2-{I'}^2)^2]/|p_n|A ,
\end{eqnarray}
where $A$ is defined as
\begin{eqnarray}
A= 1+ 2\pi^2N^2(I^2+{I'}^2)+\pi^4N^4(I^2-{I'}^2)^2.
 \label{eqn:A}
\end{eqnarray}
We will show below that the interband impurity scattering $I'$
is renormalized by $1/\sqrt{A}$.
Using eqs. (\ref{eqn:Eliash2})-(\ref{eqn:A}),
%the Eliashberg equation (\ref{eqn:Eliash2}) becomes
the frequency dependence of the gap function is obtained as
\begin{eqnarray}
\Delta(p_n) &=& 
 C\left[ Z(p_n)-\delta\Sigma^a(p_n)/\Delta(p_n) \right]^{-1}
 \nonumber \\
&=& C\left[ 1+ {2n_{\rm imp}\pi N{I'}^2}/{|p_n|A} \right]^{-1} ,
 \label{eqn:Eliash3}
\end{eqnarray}
where $C\equiv \pi N g T_{\rm c}\sum_{m}'\Delta(p_m)/|p_m|$
is a constant independent of $p_n$.
By inserting eq. (\ref{eqn:Eliash3}) into the definition of $C$,
we obtain the following equation for $T_{\rm c}$:
\begin{eqnarray}
1&=& \pi N g T_{\rm c}{\sum_m}' \left[ |p_m|
+ 2\pi n_{\rm imp}N{I'}^2/A \right]^{-1}
 \nonumber \\
&=& gN\left[ \ln \frac{\w_c}{2\pi T_{\rm c}}
-\psi\left(\frac12 + \frac{n_{\rm imp} N {I'}^2}{T_{\rm c}A}\right) \right] ,
\label{eqn:digamma}
\end{eqnarray}
where $\psi(x)$ is the digamma function.
The equation for $T_{\rm c}^0$ is given by dropping 
the term $n_{\rm imp}N{I'}^2/T_{\rm c}A$ in eq. (\ref{eqn:digamma}).
It is noteworthy that this term also vanishes 
when $x\ne1$ and $I\rightarrow\infty$, even if $n_{\rm imp}>0$.
This fact means that superconducting state is unaffected by impurities
in the unitary limit.

\begin{figure}[!htb]
\includegraphics[width=.8\linewidth]{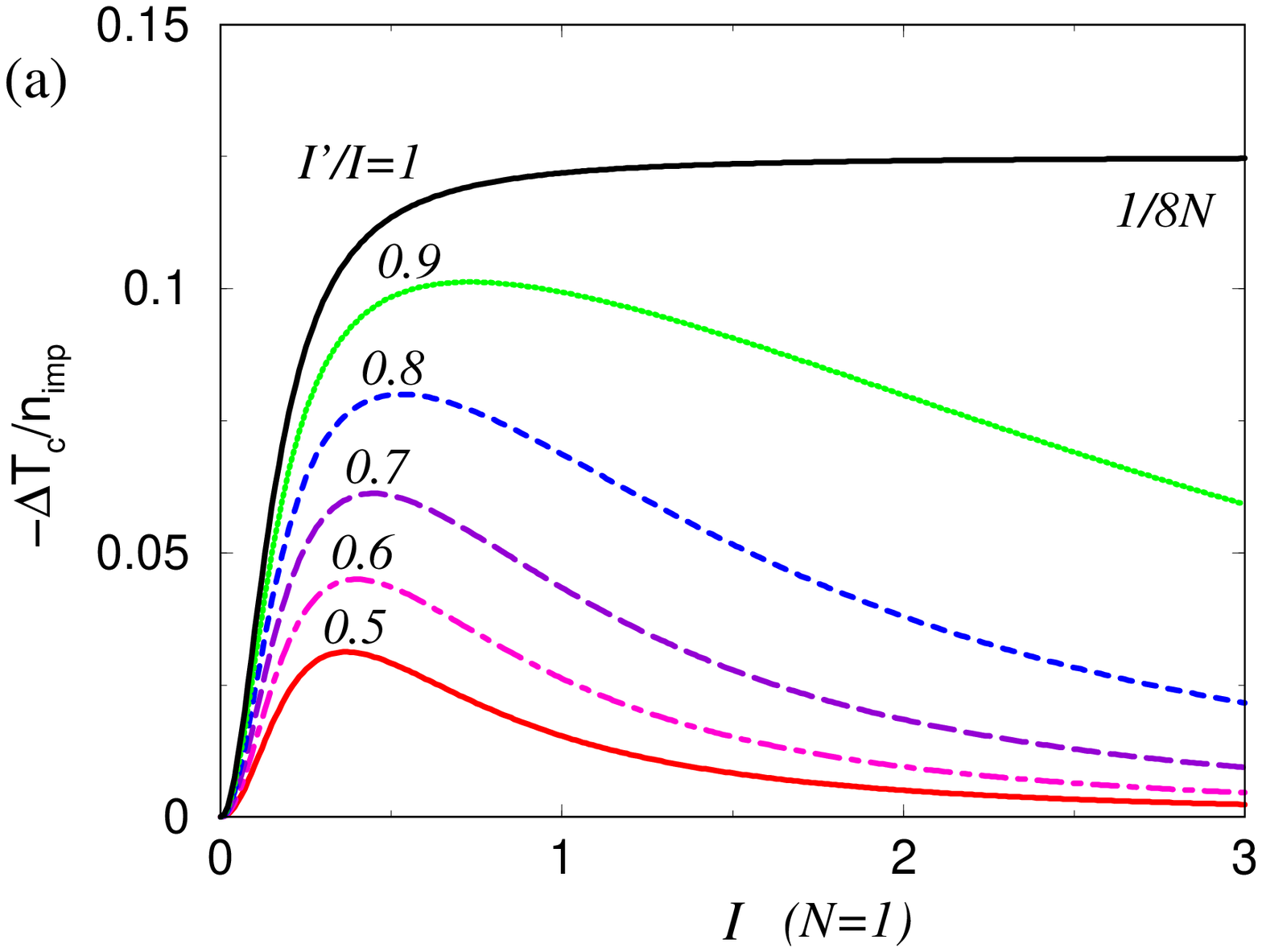}
\includegraphics[width=.8\linewidth]{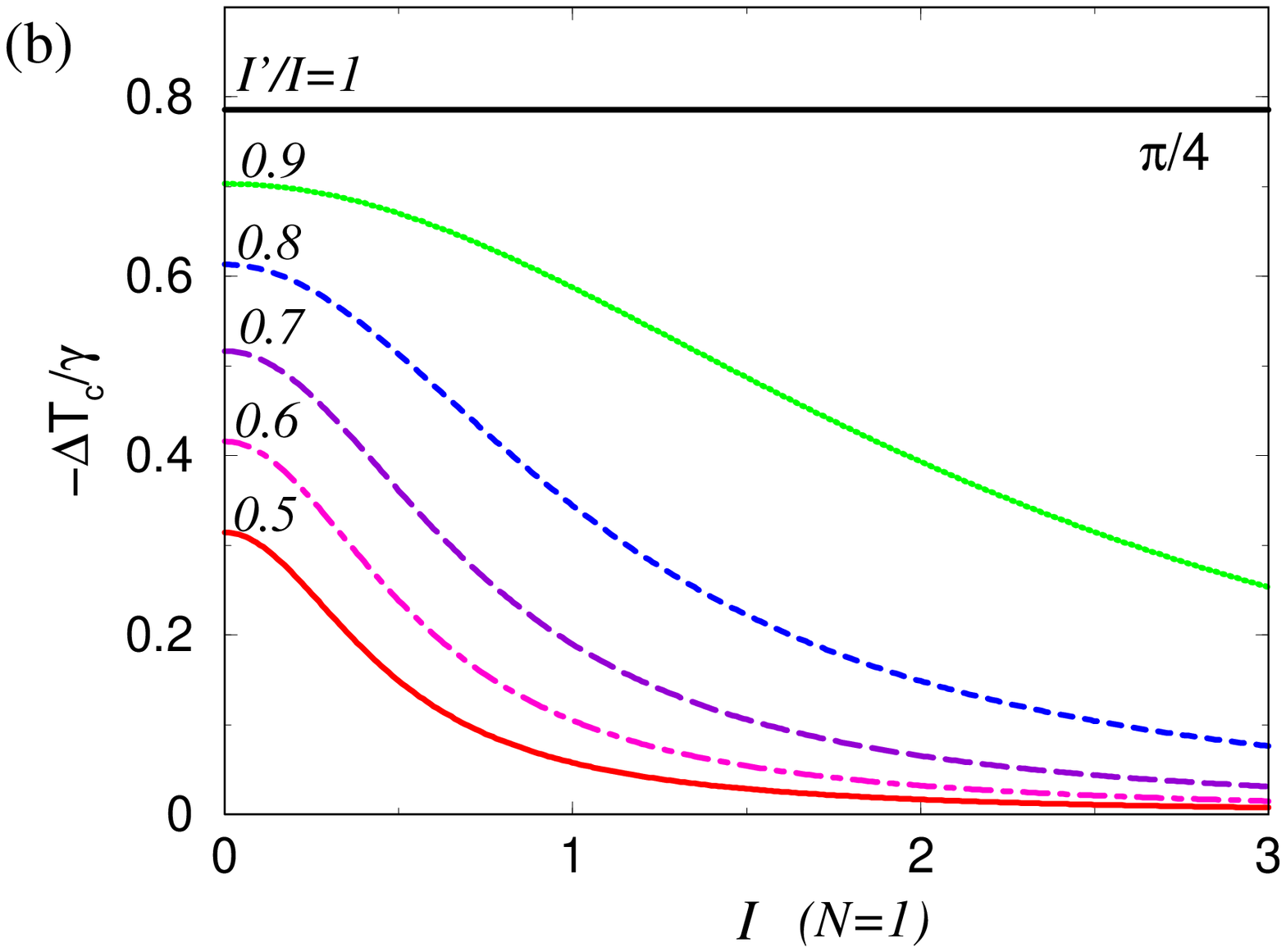}
\caption{
(a) $-\Delta T_{\rm c}/n_{\rm imp}$ and (b) $-\Delta T_{\rm c}/\gamma$
as functions of $I$.
$1/8N\sim 2500$ K in FeAs.
When $x<0.8$, $-\Delta T_{\rm c}$ becomes very small for $I\gtrsim3$.
}
\label{fig:DTc}
\end{figure}

Using eq. (\ref{eqn:digamma}) and
the relation $\psi'(1/2)= \pi^2/2$, we obtain the relationship
$\ln (T_{\rm c}^0/T_{\rm c})= n_{\rm imp}\pi^2 N {I'}^2/2T_{\rm c}^0A$
for $n_{\rm imp}\ll1$.
As a result, the reduction in $T_{\rm c}$ per impurity concentration 
for $n_{\rm imp}\ll1$ is obtained as
\begin{eqnarray}
-\frac{\Delta T_{\rm c}}{n_{\rm imp}}= \frac{\pi^2N{I'}^2}{2A}
 \label{eqn:DTcn}.
\end{eqnarray}
The physical meaning of the right-hand side of eq. (\ref{eqn:DTcn})
is the rate of pair breaking, which is given by the amplitude of 
interband scattering for Cooper pairs: $|T_{\a\b}|^2$.
Its $I$ dependence is shown in Fig. \ref{fig:DTc} (a).
When $x=1$, eq. (\ref{eqn:DTcn}) increases 
in proportion to $I^2$ in the Born regime ($\pi IN \ll 1$), 
and it approaches $1/8N$ in the unitary regime ($\pi IN \gg 1$).
In the latter case, the superconductivity in FeAs will be 
destroyed only at $n_{\rm imp}\approx 8NT_{\rm c}^0\sim 0.02$ 
since the average between 
the electron and hole density of states per Fe atom is 0.66 eV$^{-1}$ 
\cite{Singh}.
%Since $1/8N\sim 2500$ K in FeAs, the superconductivity will be 
%destroyed only at $n_{\rm imp}=0.02$ in the latter case.
When $x\ne1$, in contrast, 
eq. (\ref{eqn:DTcn}) $\approx x^2/2\pi^2N^3I^2(1-x^2)^2 \rightarrow0$
in the unitary limit.
In this case, pair breaking is almost absent
and $T_{\rm c}$ is unchanged.

%The physical meaning of the right-hand-side of eq. (\ref{eqn:DTcn})
%is the rate of pair breaking, which is given by the amplitude of 
%interband scattering for Cooper pairs; $|T_{\a\b}|^2$. % in the present model.
%It is proportional to ${I'}^2$ in the lowest order, whereas
%it is renormalized as ${I'}^2/A \sim {I'}^2\cdot(IN)^{-4} \ll {I'}^2$
%in the unitary regime for $x<1$.
%In the latter case, pair breaking is almost absent
%and $T_{\rm c}$ is unchanged.

% {\it since the $T$-matrix in 
%eq. (\ref{eqn:T}) becomes diagonal in the unitary limit as 
%$T_{\a\a}\sim iN^{-1}$ and $T_{\a\b}\sim I'(IN)^{-2} \ll I'$ except for $x=1$.}
%the $T$-matrix is renormalized as I'/[(x^2-1)(\pi IN)^2]$aaaaaa
%as the result of the multiple scattering.
%the amplitude of interband impurity scattering for Cooper pairs is 
%proportional to ${I'}^2/A \equiv {I'_{\rm eff}}^2$,
%which is equal to ${I'}^2$ in the Born limit whereas
%${I'_{\rm eff}}^2\sim {I'}^2/(IN)^4 \ll {I'}^2$ in the unitary regime.
%Thus, strong impurity potential does not cause the 
%pair breaking in fully gapped sign-reversing pairing state for $x<1$.
%since the interband scattering is strongly renormalized for $x<1$.

%%%%%%%%%%%%%%%%%%%%%%%%%%%%%%%%%
%%%%%%%%%%%%%%%%%%%%%%%%%%%%%%%%%s(

According to eqs. (\ref{eqn:DTcn}) and (\ref{eqn:gamma}),
we obtain as
\begin{eqnarray}
-\frac{\Delta T_{\rm c}}{\gamma}= 
\frac{\pi{x}^2}{2(1+{x}^2)+2\pi^2N^2I^2(1-{x}^2)^2} .
%\frac{\pi{I'}^2}{2(I^2+{I'}^2)+2\pi^2N^2(I^2-{I'}^2)^2}.
 \label{eqn:DTcg}
\end{eqnarray}
In the Born limit, $-\Delta T_{\rm c}/\gamma= \pi x^2/2(1+x^2)$,
which diminishes slowly as $x$ decreases from unity.
%approaches zero for $x\ll1$ in proportion to $x^2$.
In the unitary limit,
%$-\Delta T_{\rm c}/\gamma= \pi/4$ for $x=1$, whereas
it is strongly suppressed as
$-\Delta T_{\rm c}/\gamma \sim [-\Delta T_{\rm c}/\gamma]_{\rm Born}
\times (\pi NI)^{-1} \rightarrow 0$ for $x\ne1$.
%which approaches zero in proportion to $(IN)^{-2}$.
Note that $\gamma$ is related to residual resistivity as
$\rho_0= 2m\gamma/e^2n$, where $n$ is the carrier density.
Figure \ref{fig:DTc} (b) shows the $I$ dependence of 
eq. (\ref{eqn:DTcg}):
For $x=0.5$, $-\Delta T_{\rm c}/\gamma=0.0077$ at $IN=3$,
which is two orders of magnitude smaller than $\pi/4$.
Such a small impurity effect will be difficult to observe experimentally.

Next, we discuss the case of $N_\a\ne N_\b$, where the relation
$\Delta_\a=-\Delta_\b$ is not satisfied.
Even in this case, we can obtain $-\Delta T_{\rm c}/n_{\rm imp}$ 
by solving eq. (\ref{eqn:Eliash2}) analytically.
After a long calculation, the obtained result 
for $n_{\rm imp}\ll1$ is
\begin{eqnarray}
-\frac{\Delta T_{\rm c}}{n_{\rm imp}}= 
\frac{\pi^2\left[ \ 3(N_\a+N_\b)-2\sqrt{N_\a N_\b} \ \right]{I'}^2}
{8{\bar A}} ,
\label{eqn:DTcn2}
\end{eqnarray}
where 
${\bar A}= 1+\pi^2I^2(N_\a^2+N_\b^2)+2N_\a N_\b\pi^2{I'}^2
 + N_\a^2N_\b^2\pi^4(I^2-{I'}^2)^2$,
which is proportional to $I^4(1-x^2)^2$ in the unitary regime.
%where $\pi IN_{\a,\b}\gg1$.
Therefore, 
eq. (\ref{eqn:DTcn2}) approaches zero in the case of $x\ne1$
in the unitary regime.
Thus, the obtained results for $N_\a=N_\b$ given in 
Fig. \ref{fig:DTc} are qualitatively unchanged even for $N_\a\ne N_\b$.

%%%%%%%%%%%%%%%%%%%%%%%%%%%%%%%%%
%%%%%%%%%%%%%%%%%%%%%%%%%%%%%%%%%

On the basis of the above results,
we discuss the impurity effect on $T_{\rm c}$ in real FeAs superconductors.
Figure \ref{fig:FS} shows the Fermi surfaces (FS1-FS4) of this compound
in the unfolded Brillouin zone \cite{Kuroki,Mazin}:
%As shown in Fig. \ref{fig:FS},
%Therein, 
FS1,2 are mainly composed of $d_{xz},d_{yz}$ orbitals of Fe, whereas FS3,4 
are composed of $d_{xz},d_{yz}$ and $d_{x^2\mbox{-}y^2}$ orbitals, 
according to the five-orbital model in ref. \cite{Kuroki}.
More precisely, FS3a,4a (3b,4b) are composed of 
$d_{xz},d_{yz}$ orbitals ($d_{x^2\mbox{-}y^2}$ orbital).
Mazin et al. had proposed that superconducting gap functions in FS1,2
($\Delta_{1,2}$) and those in FS3,4 ($\Delta_{3,4}$)
are different in sign.
In this case, bands $\a$ and $\b$ in the present study
correspond to FS1,2 and FS3,4 in FeAs, respectively.

Here, we consider an impurity $d$-atom (such as Co, Ni, or Zn) 
placed at an Fe site.
In the $d$ orbital representation, the local impurity potential
will be diagonal with respect to the $d$ orbital $d(d')$:
$({\hat I})_{d,d'}= I\delta_{d,d'}$.
In the band-diagonal representation, it is transformed into
\begin{eqnarray}
I_{ij} &\approx& I \langle \sum_d 
O_{d,i}(\k) O_{d,j}(\k') \rangle^{\rm FS}_{\k\in\a,\k'\in\b} ,
% \nonumber \\
%&=&I \langle \sum_d \langle \a;\k|d;\k \rangle \langle d;\k'|\b;\k'\rangle 
%\rangle^{\rm FS}_{\k\in\a,\k'\in\b} ,
\end{eqnarray}
where $i,j\ (=\a,\b)$ represent the band indices.
$O_{d,i}(\k)\equiv \langle d;\k|i;\k\rangle$
is the transformation orthogonal (or unitary) matrix 
between the orbital representation (orbital $d$) and the 
band-diagonal representation (band $i$).
Therefore, $I_{\a\a},\ I_{\b\b}\approx I$
whereas $|I_{\a\b}|$ should be smaller than $|I|$
when bands $\a$ and $\b$ are composed of different $d$ orbitals.
Since FS3b,4b are composed of $d_{x^2\mbox{-}y^2}$ orbitals of Fe,
the relation $x=|I'/I|\sim 0.5$ is realized in FeAs.

On the other hand, according to the RPA analysis
by Kuroki et al. \cite{Kuroki}, $\Delta_{3,4}$ has line nodes
near $(\pi,0)-(0,\pi)$ line
owing to spin fluctuations with $\q\approx(\pi,\pi/2)$;
therein, the sign of $\Delta_{3,4}$ for FS3a,4a (3b,4b) and 
that of $\Delta_{1,2}$ are equal (different).
Then, bands $\a$ and $\b$ in the present study
correspond to [FS1,2+3a,4a] and [FS3b,4b] in FeAs, respectively.
Since bands $\a$ and $\b$ are composed of different $d$ orbitals,
the relation $x\ll1$ is expected.
Note that the obtained $s$-wave state 
is fully gapped since the line nodes on FS3,4, which are not protected 
by symmetry, are masked by small interband pairing \cite{Kuroki}.
We also note that the sign change in $\Delta_{3,4}$ does not
occur and Mazin's type sign-reversing state is realized 
in the case of $U\approx U'$ \cite{Kuroki}.

As a result, in both unconventional $s$-wave states
proposed in refs. \cite{Kuroki} and \cite{Mazin},
$-\Delta T_{\rm c}/\gamma^{\rm imp}\approx0$ in the unitary regime.
In high-$T_{\rm c}$ cuprates, $I$ due to Zn impurity is 
about 10 eV \cite{Kontani-review}.
If we expect $I\sim10$ eV and $N_{\a,\b}\sim1$ eV$^{-1}$ in FeAs,
the unitary regime is actually achieved.
%Thus, the smallness of impurity effect in FeAs superconductors 
%does not contradicts with these theoretical predictions.

Although we have put $g_{\a\a}=g_{\b\b}=0$ above,
they will take positive (negative) values owing to electron-electron 
correlation (electron-phonon interaction) in FeAs superconductors.
Although these diagonal interactions modify $T_{\rm c}^0$,
we can show that both eqs. (\ref{eqn:DTcn}) and (\ref{eqn:DTcg}) 
are unchanged even if $g_{\a\a},g_{\b\b}\ne0$.
Thus, the obtained results in the present study will be valid for 
general multiband fully gapped superconductors with sign reversal.

In summary, we analyzed the impurity effect on $T_{\rm c}$
in a sign-reversing $s$-wave BCS model.
%based on the $T$-matrix approximation.
%In the case of $x=I'/I=1$, $-\Delta T_{\rm c}$ is large 
%due to the interband transition of Cooper pairs induced by $I'$.
Except at $I'=I$,
$T_{\rm c}$ is almost unaffected by impurities in the unitary limit,
since the interband elements of the $T$-matrix vanish due to
multiple scattering.
Thus, the robustness of superconductivity against Co, Ni, or Zn impurities
in this compound is naturally explained 
in terms of the sign-reversing $s$-wave superconductivity 
 \cite{Kuroki,Mazin,RG,Nomura,Yanagi,Tesanovic}. 
On the other hand, a weak impurity scattering causes pair breaking.
Therefore, the finite density of states 
may be induced in the superconducting state by weak impurities or disorder.
%The reduction in $T_{\rm c}$ is large when $x=I'/I=1$.
%In highly contrast, it quickly approaches zero as $x$ decreases from unity.
%Especially, $T_{\rm c}$ is almost unaffected by impurities 
%in the unitary regime as shown in Fig. \ref{fig:DTc}
%since the $T$-matrix becomes diagonal in the unitary limit,
%which means that the pair breaking due to interband scattering 
%is absent.
%suppressed as $T_{\a\b} \propto I'\cdot (IN)^{-2}\ll I'$
%by the factor $1/(\pi IN)^2 \ll1$
%due to multiple scattering. % in multiband system.
%In FeAs superconductors $x<1$ is satisfied since the hole pockets 
%and electron pockets are not composed of the same $d$ orbitals.

We are grateful to M. Sato and D.S. Hirashima for 
stimulating daily discussions.
We are also grateful to Y. Matsuda, T. Shibauchi,
H. Aoki, K. Kuroki, R. Arita, Y. Tanaka, S. Onari and 
Y. ${\bar {\rm O}}$no for useful comments and discussions.
This study has been supported by Grants-in-Aids for Scientific
Research from MEXT, Japan.

%%%%%%%%%%%%%%%%%%%%%%%%
%references
%%%%%%%%%%%%%%%%%%%%%%%%

\end{document}